\documentclass[aps,prb,reprint,superscriptaddress,nobibnotes]{revtex4-1}

\usepackage{siunitx}
\usepackage{graphicx}
\usepackage{hyperref}
\usepackage{amsmath}
\usepackage{bm}
\usepackage{xr}
\usepackage{float}

\begin{document}


\newcommand{\shcunit}{($\hbar/e$)($\Omega$cm)\textsuperscript{-1}} 
\newcommand{\ef}{Fermi energy}

\title{Intrinsic Spin Hall Conductivity in MoTe\textsubscript{2} and WTe\textsubscript{2} Semimetals}

\author{Jiaqi Zhou}
\affiliation{Fert Beijing Institute, BDBC, School of Electronic and Information Engineering, Beihang University, Beijing 100191, China}
\affiliation{Centre de Nanosciences et de Nanotechnologies, CNRS, Universit\'{e} Paris-Sud, Universit\'{e} Paris-Saclay, Orsay 91405, France}

\author{Junfeng Qiao}
\affiliation{Fert Beijing Institute, BDBC, School of Electronic and Information Engineering, Beihang University, Beijing 100191, China}

\author{Arnaud Bournel}
\affiliation{Centre de Nanosciences et de Nanotechnologies, CNRS, Universit\'{e} Paris-Sud, Universit\'{e} Paris-Saclay, Orsay 91405, France}

\author{Weisheng Zhao}
\email{weisheng.zhao@buaa.edu.cn}
\affiliation{Fert Beijing Institute, BDBC, School of Electronic and Information Engineering, Beihang University, Beijing 100191, China}

\date{\today}

\begin{abstract}

We report a comprehensive study on the intrinsic spin Hall conductivity (SHC) in semimetals MoTe\textsubscript{2} and WTe\textsubscript{2} by the \textit{ab initio} calculation. Large SHC and desirable spin Hall angles have been discovered, due to the strong spin orbit coupling effect and low charge conductivity in semimetals. Diverse anisotropic SHC values, attributed to the unusual reduced-symmetry crystalline structure, have been revealed. 
 We report an effective method on SHC optimization by electron doping, and exhibit the mechanism on SHC variation with energy shifting by the spin Berry curvature. Our work provides insights into the realization of strong spin Hall effects in 2D systems. 

\end{abstract}


\maketitle

\section{introduction}

Spin-orbit torque (SOT) generated by materials with strong spin-orbit coupling (SOC) is a promising approach for energy-efficient manipulation of nonvolatile magnetic memory and spin logic devices\cite{Baek2018Mar,Zhang2018Apr,Shi2018Jan,science2012,Manchon2009Mar}. 
The form of the in-plane SOT alone does not allow for deterministic switching of perpendicular magnetic anisotropy (PMA) devices, and magnetization switching requires an additional external in-plane magnetic field\cite{Liu2012Aug,Han2017Aug}. 
Efforts are devoted to find efficient methods on PMA devices switching by field-free SOT.
It is found that lateral structural asymmetry, the wedge film, makes it possible to switch the PMA device by in-plane-current SOT without external magnetic fields\cite{Yu2014May}. Besides, the out-of-plane SOT can be introduced by WTe\textsubscript{2}, a layered orthorhombic transition-metal dichalcogenide (TMD) with strong SOC and reduced crystalline symmetry\cite{MacNeill2016Nov,MacNeill2017Aug}.
WTe\textsubscript{2}/Py device produces an out-of-plane antidamping torque when current is applied along a low-symmetry axis, but not when current is applied along a high-symmetry axis. This is due to the absence of two-fold rotational symmetry in WTe\textsubscript{2}/Py bilayer.The anisotropic enhancement of spin-orbit torques in WTe\textsubscript{2}/Py devices has been observed\cite{Li2018Sep}. 

On the other hand, one SOT mechanism is the bulk spin Hall effect (SHE). A charge current flowing in the spin Hall layer can generate a pure spin current that exerts a spin torque on the recording layer\cite{Khang2018Jul}. The SHE can be separated into intrinsic and extrinsic parts. The intrinsic SHE, significantly contributing to the total SHE in materials with strongly spin-orbit-coupled bands, can be calculated accurately based on \textit{ab initio} theories\cite{Qiao2018Dec,Guo2008,Sui2017Dec}. As the materials with broken crystal symmetry provide anisotropic SOT\cite{MacNeill2016Nov,Li2018Sep}, it triggers an intriguing question on the intrinsic SHE of the low-symmetry crystals. Exploring the intrinsic SHE in low-symmetry crystal would greatly contribute to research on novel materials with diverse anisotropy and applications. Besides, it is expected to discover high spin-charge conversion efficiency in semimetals considering their low conductivity.

In this work, we report the SHC in semimetals MoTe\textsubscript{2} and WTe\textsubscript{2} using \textit{ab initio} calculations. These studies reveal large SHC and high spin Hall angle in MoTe\textsubscript{2} and WTe\textsubscript{2}, as well as diverse anisotropic SHC values. We analyze the mechanism on SHC by spin Berry curvature, and illustrate that energy shifting is an effective method to enhance SHC.


\section{METHOD}


The general form of Kubo formula SHC is given by 
\cite{Yao2004,Guo2005,Matthes2016} 
\begin{align}
\begin{split}
\label{equ:kubo}
\sigma_{xy}(\omega) = & \frac{\hbar}{V N_k^3}
\sum_{\bm{k}}\sum_{n} f_{n\bm{k}} \\
& \times \sum_{m \neq n}
\frac{2\operatorname{Im}[\langle n\bm{k}| \hat{j}_{x}|m\bm{k}\rangle
	\langle m\bm{k}| -e\hat{v}_{y}|n\bm{k}\rangle]}
{(\epsilon_{n\bm{k}}-\epsilon_{m\bm{k}})^2-(\hbar\omega +i\eta)^2},
\end{split}
\end{align}
where $\hat{j}_{x}=\frac{1}{2}\{\hat{s}_z,\hat{v}_x\}$, $V$ is the primitive cell 
volume, and $N_k^3$ is the number of $k$-points in the Brillouin zone (BZ). 
To facilitate further analysis, we separate the 
Equ.(\ref{equ:kubo}) into the band-projected 
Berry curvature-like term
\begin{align}
\begin{split}
\label{equ:kubo_shc_berry}
\Omega_{n,xy}^{\text{spin}z}(\bm{k}) = {\hbar}^2 \sum_{
	m\ne n}\frac{-2\operatorname{Im}[\langle n\bm{k}| 
	\frac{1}{2}\{\hat{\sigma}_z,\hat{v}_x\}|m\bm{k}\rangle
	\langle m\bm{k}| \hat{v}_{y}|n\bm{k}\rangle]}
{(\epsilon_{n\bm{k}}-\epsilon_{m\bm{k}})^2-(\hbar\omega+i\eta)^2}
\end{split}
\end{align}
and $k$-resolved term
\begin{equation}
\label{equ:kubo_shc_berry_sum}
\Omega_{xy}^{\text{spin}z}(\bm{k}) = \sum_{n}
f_{n\bm{k}} \Omega_{n,xy}^{\text{spin}z}(\bm{k}).
\end{equation}
The SHC is the sum over occupied bands
\begin{equation}
\label{equ:kubo_shc_sum}
\sigma_{xy}^{\text{spin}z}(\omega) = 
-\frac{e^2}{\hbar}\frac{1}{V N_k^3}\sum_{\bm{k}}
\Omega_{xy}^{\text{spin}z}(\bm{k}).
\end{equation}

All the \textit{ab initio} calculations were performed using 
{\sc Quantum ESPRESSO} package based on projector-augmented wave (PAW) 
method and a plane wave basis set \cite{Giannozzi2009,Giannozzi2017Oct}. 
The exchange and correlation terms were described using 
generalized gradient approximation (GGA) in the scheme of 
Perdew-Burke-Ernzerhof (PBE) parameterization, as implemented in the 
{\sc pslibrary} \cite{Corso2014}. The plane-wave cutoff energy is 600 eV and a $k$-point grid with $12\times10\times6$ was used in self-consistent calculation. We employed the atomic structure in previous reports\cite{Wang2016Jul,Soluyanov2015Nov}. Then, density functional theory (DFT) wave functions were projected to maximally localized Wannier functions using the WANNIER90 package \cite{Marzari1997,Souza2001,Marzari2012} and the Kubo formula was employed to calculate the SHC. A dense $k$ mesh of 500$\times$500$\times$500 was employed to perform the BZ integration for the intrinsic SHC, and adapt $k$ mesh was used to deal with the dramatic variation in the spin Berry curvature. More details can be found in our previous work\cite{Qiao2018Dec}.

\section{RESULTS AND DISCUSSIONS}

\begin{figure}[tb]
	\includegraphics{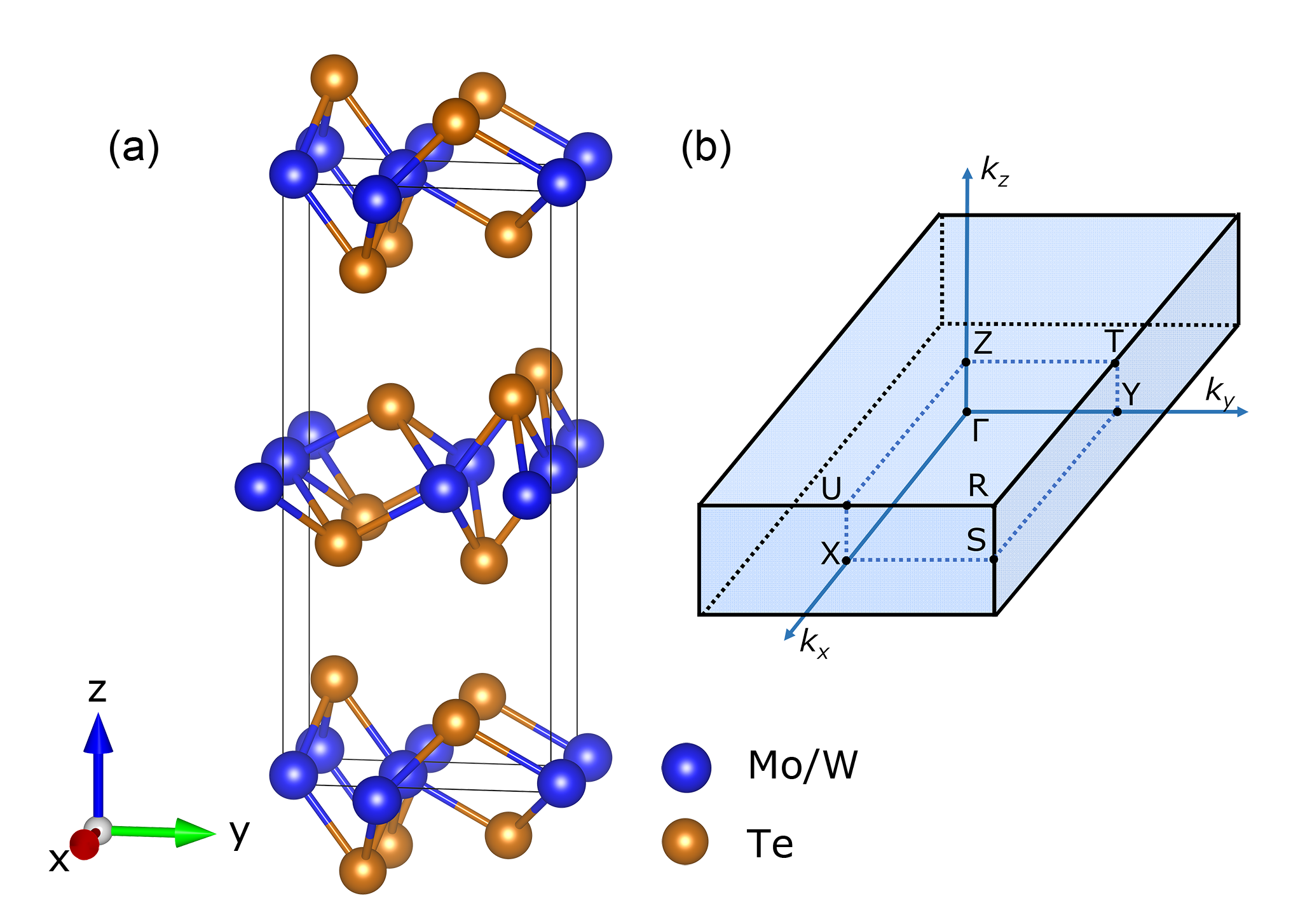}
	\caption{\label{fig:model}
		(a) Three-dimensional view of the orthorhombic crystal structure of T\textsubscript{d}-MTe\textsubscript{2} (M=Mo or W) in the space group of Pnm2\textsubscript{1}.The blue and tawny balls represent M and Te atoms, respectively. (b) Schematic diagram of the three-dimensional BZ of MTe\textsubscript{2}, the high symmetry points are indicated by black dots.}
\end{figure}

 Figure \ref{fig:model}(a) shows the atomic structure of MTe\textsubscript{2} (M = Mo or W) semimetal. Both MoTe\textsubscript{2} and WTe\textsubscript{2} are layered orthorhombic transition-metal dichalcogenides (TMD) with strong SOC\cite{Sun2015Oct,Feng2016Nov}. The space group is identified to be Pnm2\textsubscript{1} (No. 31). The crystal structure possesses one mirror plane $x$ = 0 and one glide mirror plane parallel to $y$ = 0, which transform to $k_x$ = 0 and $k_y$ = 0 plane, respectively, in the momentum space.
 

\begin{table}[t]
	\caption{\label{tab:shc}
		The calculated non-zero SHC tensor elements in MoTe\textsubscript{2} and WTe\textsubscript{2} at Fermi energy, in unit of ($\hbar/e$) ($\Omega$cm)\textsuperscript{-1} The maximum SHA values are listed in the last column.
		}
	\begin{ruledtabular}
		\begin{tabular}{c c c c c c c c}
Semimetals	        &  $\sigma_{yz}^{x}$     & $\sigma_{zy}^{x}$  & $\sigma_{xz}^{y}$   & $\sigma_{zx}^{y}$        & $\sigma_{xy}^{z}$       & $\sigma_{yx}^{z}$     & SHA\textsubscript{MAX} \\                                                			                                                    
\hline
MoTe\textsubscript{2}  & \num{-18}		 & \num{45}	      & \num{-88}	    & \num{286}	               & \num{-176}		 & \num{-361}	           & -0.72\\
WTe\textsubscript{2}  & \num{-44}		 & \num{-44}	      & \num{-61}	    & \num{103}		       & \num{-204}		 & \num{-15}	           & -0.54\\
		\end{tabular}
	\end{ruledtabular}
\end{table}

\begin{figure}[b]
	\includegraphics{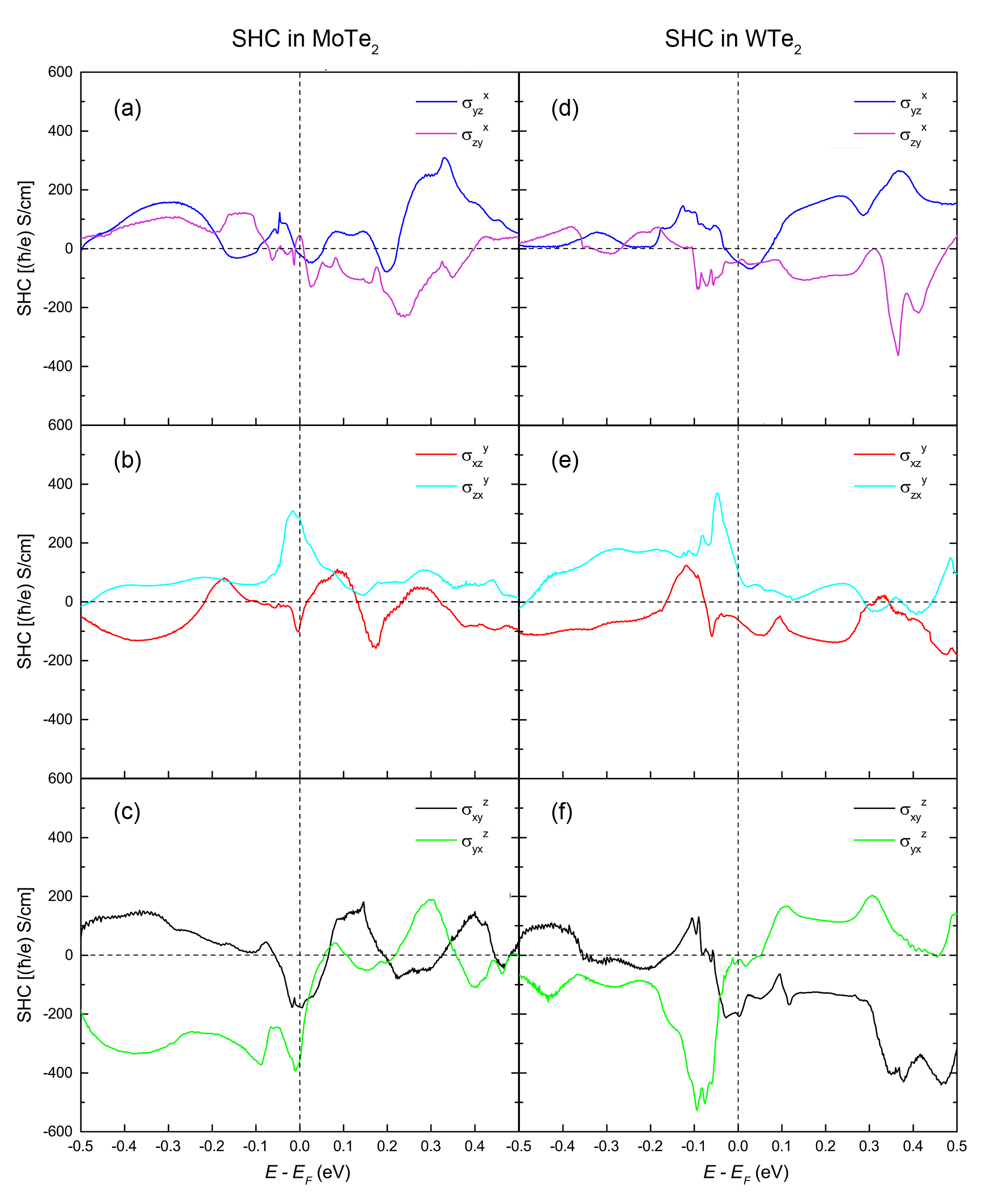}
	\caption{\label{fig:shc}
		Energy dependent SHC tensor elements for (a)(b)(c) MoTe\textsubscript{2} and (d)(e)(f)  WTe\textsubscript{2}. Lines in different colors represent different kinds of SHC. The dashed vertical line corresponds to the Fermi energy and the dashed horizontal line represents the SHC is zero.}
\end{figure}

Table \ref{tab:shc} shows the calculated SHC for MoTe\textsubscript{2} and WTe\textsubscript{2} at Fermi energy.  Common metals exhibiting $C_{4v}$ symmetry such as Pt, W, Ta, have the limitation on $\sigma_{xy}^{z}$ = $-\sigma_{yx}^{z}$, $\sigma_{yz}^{x}$ = $-\sigma_{xz}^{y}$, and $\sigma_{zx}^{y}$ = $-\sigma_{zy}^{x}$.
$\beta-$Ta and $\beta-$W have more strict restrictions as $\sigma_{xy}^{z}$ = $-\sigma_{yx}^{z}$
 = $\sigma_{yz}^{x}$ 
 = $-\sigma_{zy}^{x}$
 = $\sigma_{zx}^{y}$
 = $-\sigma_{xz}^{y}$. But in MoTe\textsubscript{2} and WTe\textsubscript{2}, we found more diverse anisotropies than heavy metals.
   Both MoTe\textsubscript{2} and WTe\textsubscript{2} have strong anisotropies with six non-zero SHC tensors with different absolute values, which vary a lot from positive to negative scale. For MoTe\textsubscript{2}, the SHC max is $\sigma_{yx}^{z}$ = -360\shcunit, and $\sigma_{zx}^{y}$ = 286\shcunit is also a considerable one. For WTe\textsubscript{2}, the maximum of SHC is $\sigma_{xy}^{z}$ =-204\shcunit. These results will provide helpful information for the experimental detection of the SHE. 
     We also analyzed the spin Hall angle (SHA). SHA is the SHE efficiency and the critical current density for magnetization switching, which measures the efficiency of the charge current to spin current conversion\cite{Zhang2016Sep}. The SHA\textsubscript{MAX} of MoTe\textsubscript{2} is -0.72 and the SHA\textsubscript{MAX} of WTe\textsubscript{2} is -0.54.
We take Pt to make a comparison. Pt is a 5$d$ heavy metal with the SHC in the order of 10\textsuperscript{3} and the conductivity in the range of 10\textsuperscript{4}--10\textsuperscript{6} ($\Omega$cm)\textsuperscript{-1}\cite{Guo2008}, consequently the spin Hall angle is relatively small\cite{Wang2014Oct, Zhang2015Apr}. On the other hand, the conductivities of these two semimetals are much lower.  The conductivity of WTe\textsubscript{2} is  7.4$\times$10\textsuperscript{2}($\Omega$cm)\textsuperscript{-1}\cite{Jana2015Jun} while the conductivity of MoTe2 is 1$\times$10\textsuperscript{3}($\Omega$cm)\textsuperscript{-1} \cite{Qi2016Mar}. As a result, high SHA can be expected in MoTe\textsubscript{2}  and WTe\textsubscript{2}   as shown in Tab. \ref{tab:shc}.

According to the Kubo formula, SHC varies quickly with the Fermi energy. Figure \ref{fig:shc} presents the variation of SHC with respect to the position of Fermi energy. In the left panel of Fig.\ref{fig:shc}, MoTe\textsubscript{2} shows peaks around Fermi energy for $\sigma_{yx}^{z}$, $\sigma_{zx}^{y}$ and $\sigma_{xy}^{z}$, as high as -390, 306 and -178\shcunit, respectively. The relatively high SHC, over 200\shcunit, remains from -0.5 eV to 0 eV for $\sigma_{yx}^{z}$. In the right panel of Fig. \ref{fig:shc}, WTe\textsubscript{2} exhibits SHC as high as $\sigma_{yx}^{z}$ = -528\shcunit when the Fermi energy lies at -0.094 eV, the high SHC around -400\shcunit, arises at 0.47 for $\sigma_{xy}^{z}$, and the $\sigma_{zx}^{y}$ value reaches up to 370 when the Fermi energy lies at -0.048 eV.  The energy-dependent analysis indicates the general route to optimizing the SHE in these materials. Shifting Fermi energy by weak electron doping is an effective way to enhance SHC in MoTe\textsubscript{2} and WTe\textsubscript{2}.

\begin{figure}[!htb]
	\includegraphics{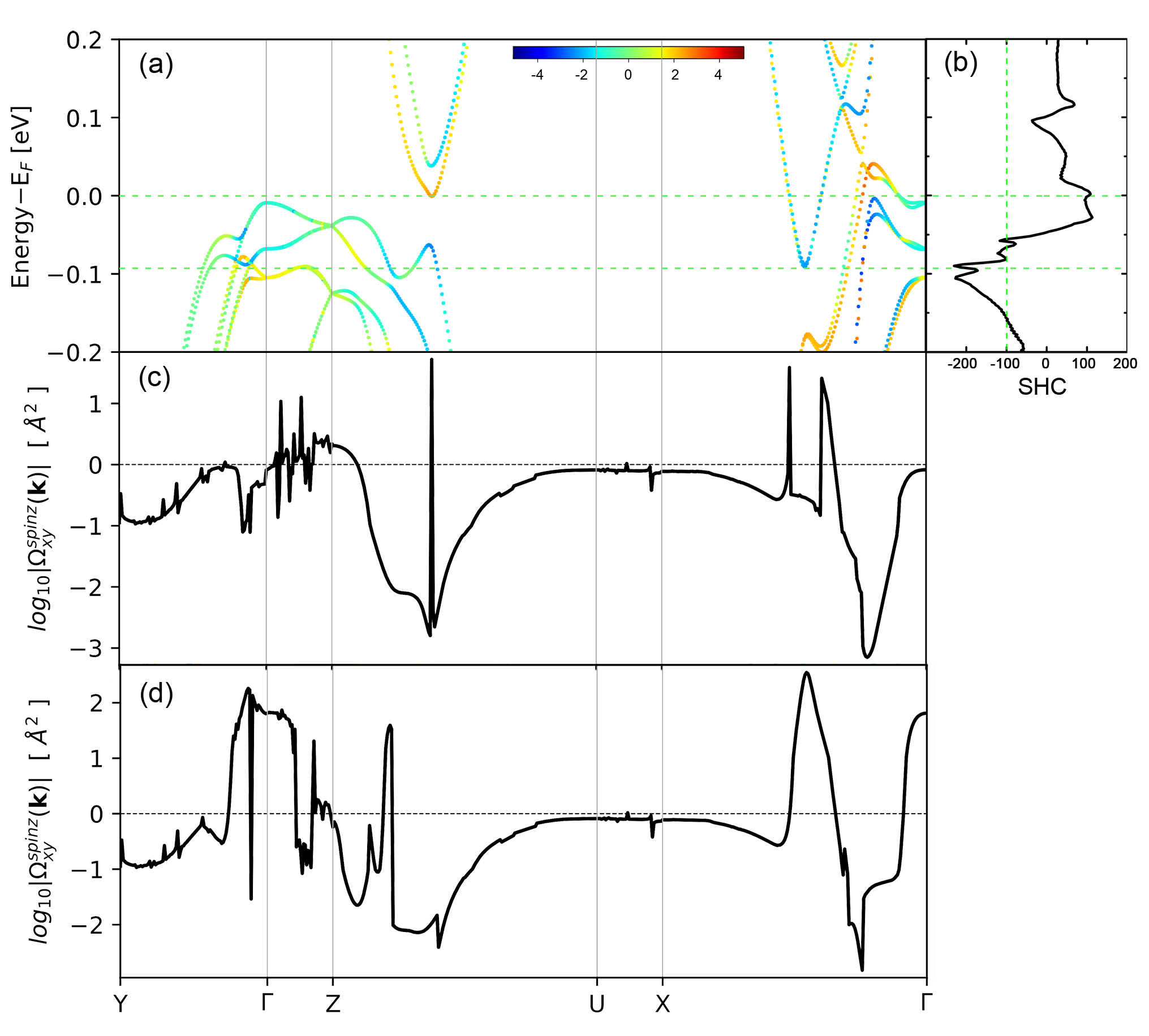}
	\caption{\label{fig:kpath}
		(a) Band-projected spin Berry curvature. (b) $\sigma_{xy}^{z}$ SHC in WTe\textsubscript{2} as a function of energy. Two green dashed horizontals crossing (a) and (b) represent $E=E_F$ and $E=E_F - 0.09$, respectively.  The $k$-resolved spin Berry curvatures at  $E=E_F$ and $E=E_F - 0.09$ are shown in (c) and (d), respectively. }
\end{figure}

To elucidate the underlying mechanism of the enhancement of SHC induced by energy shift, we performed band-projected and $k$-resolved spin Berry curvature in Figure \ref{fig:kpath}. We take $\sigma_{xy}^{z}$ in WTe\textsubscript{2} as an example, as it varies a lot and reverses in a small energy range, from -204\shcunit at $E=E_F$ to 127\shcunit at $E=E_F - 0.09$, marked by green horizontal dashed line crossing Fig. \ref{fig:kpath}(a) and (b). Fig. \ref{fig:kpath}(a) shows the $k$-resolved band projected by spin Berry curvature $\Omega_{n,xy}^{\text{spin}z}(\bm{k})$ with log scale, where red (blue) color denotes a positive (negative) contribution of spin Berry curvature. The bands close to Fermi energy mainly contribute to SHC, especially at spin-orbit splitting points. As the SHC is the sum over the occupied bands, the Fermi energy plays an important role in SHC scale.  SHC at Fermi energy presents a minus sign due to negative contribution along Z-U and X-$\Gamma$ point. When the energy is shifted by -0.09, positive spin Berry curvature rises and overwhelms the negative one, resulting in a 127\shcunit  SHC. 
To make it more visualized, Fig. \ref{fig:kpath}(c) and (d) show the $k$-resolved spin Berry curvature with log scale.  At Fermi energy, the negative spin Berry curvature dominants along the Z-U and X-$\Gamma$ path. Some sharp peaks are caused by the bands crossing with Fermi energy. When the energy is shifted by -0.09eV, the negative spin Berry curvature along the X-$\Gamma$ path is weakened while the positive one strengthens, especially at the $\Gamma$ point. As a result the SHC turns to be positive. It is remarkable that the spin Berry curvature varies dramatically  along the $k$-path. In such cases, the method of adaptive $k$ mesh refinement can be effective for refined SHC calculation.

\begin{figure}[tb]
	\includegraphics[width=0.4\textwidth]{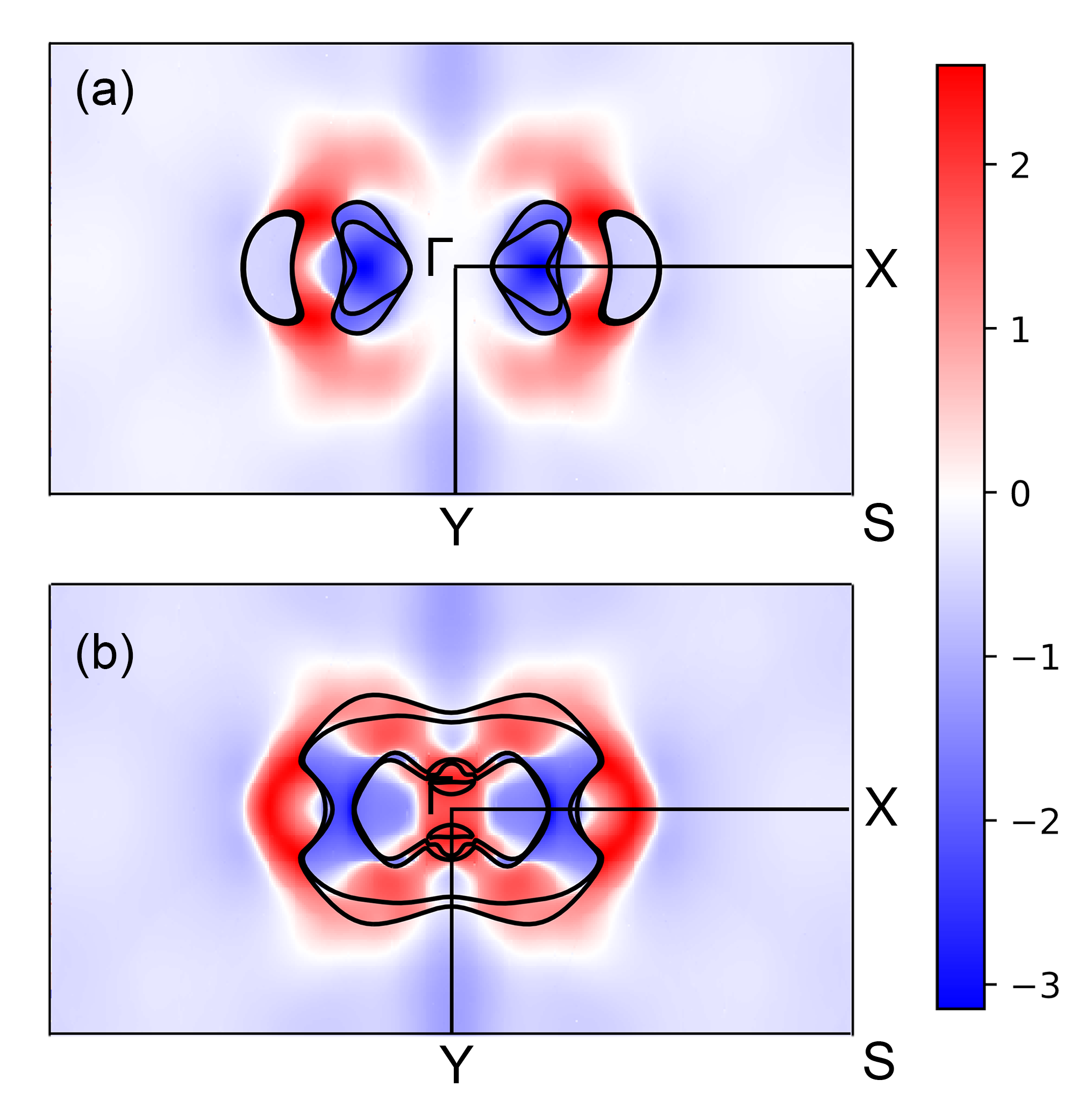}
	\caption{\label{fig:kslice}
		The $k$-resolved spin Berry curvature in the 2D BZ at $k_z$ = 0 for  $\sigma_{xy}^{z}$ SHC in WTe\textsubscript{2} at (a) $E = E_F$ and (b)  $E = E_F - 0.09$, respectively. High symmetry points are  $\Gamma$ (0, 0, 0), X (0.5, 0, 0), Y (0, 0.5, 0) and S (0.5, 0.5, 0).}
\end{figure}

We show the $k$-resolved spin Berry curvature in the two-dimensional BZ  at $k_z$ = 0 at the $E = E_F$ and $E = E_F - 0.09$ points in Figure\ref{fig:kslice}(a) and (b), respectively.  The spin Berry curvature is also in the log scale and red (blue) color denotes a positive (negative) contribution.
It can be seen that the spin Berry curvature depends sensitively on energy shift including a sign reversal throughout a large fraction of the BZ, especially around the $\Gamma$ point. Fig.\ref{fig:kslice}(a) shows that at $E = E_F$ point, the spin Berry curvature at $\Gamma$ point is close to zero, and the negative region surrounded by black line is very conspicuous considering the log scale. At $E = E_F - 0.09$ point in Fig.\ref{fig:kslice}(b), the spin Berry curvature around $\Gamma$ point becomes positive shown by red color. Besides, the positive region in red color expands widely while the negative region is weakened. As a result of the spin Berry curvature integration, the SHC inverts its sign from the negative one at $E = E_F$ to the positive one at  $E = E_F - 0.09$.
The analysis above clarifies the mechanism on SHC variation with energy, and sheds light on the SHC optimization.

\section{CONCLUSION}

In summary, by \textit{ab initio} calculations, we present the anisotropic spin Hall conductivity in MoTe\textsubscript{2} and WTe\textsubscript{2} semimetals.  The SHC is desirable in both semimetals, and the spin Hall angle is conspicuously higher compared with heavy metals. We find the reduced symmetry results in the diverse SHC anisotropies, which provides potential for various application. We report the possibility to improve SHC by energy shift, which can induce a large enhancement or reversal of the spin Berry curvature integration by doping. Our investigations are conducive to the study on the spin Hall effect and spin-orbit torque in 2D systems. 

\begin{acknowledgments}

The authors gratefully acknowledge the National Natural Science 
Foundation of China (Grant No. 61627813, 61571023), the International 
Collaboration Project B16001, and the National Key Technology Program 
of China 2017ZX01032101 for their financial support of this work. This work is supported by the Academic Excellence  Foundation of BUAA for PhD Students.

\end{acknowledgments}


%

\end{document}